\documentstyle[preprint,aps]{revtex}

\begin{document}
\topmargin -1.4cm

\draft
\title{Double quantum dots: interdot interactions, co-tunneling,
and Kondo resonances without spin} 

\author{Qing-feng Sun and Hong Guo}

\address{Center for the Physics of Materials and Department
of Physics, McGill University, Montreal, PQ, Canada H3A 2T8}

\maketitle

\begin{abstract}
We show that through an interdot off-site electron correlation in
a double quantum-dot (DQD) device, Kondo resonances emerge in 
the local density of states without the electron spin-degree of 
freedom. We identify the physical mechanism behind this phenomenon: 
rather than forming a spin singlet in the device as required 
in the conventional Kondo physics, we found that exchange of 
electron position between the two quantum dots, together with 
the off-site Coulomb interaction, are sufficient to induce 
the Kondo resonance. Due to peculiar co-tunneling events,
the Kondo resonance in one dot may be pinned at the chemical 
potential of the other one.

\end{abstract}

The recent discovery of Kondo effect in semiconductor quantum dots
(QD)\cite{ref2,ref3} has generated tremendous interests both theoretically
and experimentally. When a QD is contacted by two metallic leads, Kondo
effect arises as a result of coherent superposition of co-tunneling
processes\cite{ref3,ref1} such that, at low temperatures, a very narrow Kondo 
resonance emerges in the density of states of the QD at the free electron 
chemical potential $\mu$, and a spin singlet is formed in the whole system 
of the QD plus the leads. Kondo effect has so far been investigated in 
many systems and its understanding provided important insights to 
strongly interacting electrons in nanostructures. In most previous 
theoretical investigations, the physical systems under consideration
all possess three common and fundamental ingredients:\cite{ref5} the 
existence of co-tunneling ({\it i.e.} exchange) events, the role of the 
electron spin degrees of freedom, and finally the on-site Coulomb 
interaction $U_o$. Because the QD Kondo effect is a result of 
many-body correlation, the existence of a great deal of co-tunneling 
events and the Coulomb interaction are crucial. However, must it 
be the on-site interaction?  Are spin degrees of freedom essential? 
Although no exception has been found so far, these questions are 
intriguing because their understanding will provide a clear picture 
to the deciding physical factors of QD Kondo phenomenon. Our 
investigation found that QD Kondo effect does occur even without 
considering the spin degrees of freedom and without the on-site 
Coulomb energy --- provided that there are spatial degrees of 
freedom and there is an off-site interaction. It is the purpose 
of this letter to report our investigations on these issues.

In particular, we investigate a two-probe double QD (DQD) device 
in which the spin degeneracy is {\it neglected}. The two QDs provide a 
spatial degree of freedom for co-tunneling processes (see below) and 
there is an off-site e-e interaction between the dots. It should be mentioned 
that conventional Kondo effect, {\it i.e.} the Kondo effect caused by 
the three fundamental ingredients mentioned in the last paragraph,
in DQD systems have been the subject of several recent 
studies\cite{ref7,ref8,ref9}. These investigations considered effects of the 
intradot on-site interaction $U_o$ and/or interdot magnetic spin exchange. 
The latter is not an independent effect as it is determined by the 
(intradot) interaction $U_o$ and the interdot tunnel coupling\cite{ref9}. 
The present work, however, focuses on Kondo effect induced by the interdot 
off-site interaction $U$ without the spin degrees of freedom. 
Our main 
findings are: (i) At low temperatures there are Kondo resonances in the local 
density of states (LDOS) of the DQD which occur without the entire system
forming a spin singlet; (ii) If the interdot tunnel coupling is zero, 
the Kondo resonance in one QD is pinned at the chemical potential of the lead
that is attached to the {\it other} dot; (iii) At a non-zero interdot
tunnel coupling, this Kondo resonance is split so that sharp features
including peaks and abruptly drops emerge in the conductance of the device.
These results are significant because they show that the Kondo resonance 
can occur by an off-site interaction even without the electron 
spin-degrees of freedom.

Our system of DQD coupled with two leads is described by the following 
Hamiltonian, $H=H_{DQD}+H_T+\sum\limits_{\alpha(\alpha=L,R)}H_{\alpha}$,
where $H_{\alpha}=\sum\limits_k \epsilon_{\alpha k}a^{\dagger}_{\alpha k}
a_{\alpha k}$ describe the noninteraction left/right lead;
$H_T=\sum_{k,\alpha} (t_{\alpha}a^{\dagger}_{\alpha k}d_{\alpha} + H.c.)$ 
denotes the tunneling between the left (right) QD and the left (right) lead.
$H_{DQD}=\sum_{\alpha}\epsilon_{\alpha} d^{\dagger}_{\alpha}d_{\alpha}
+Ud^{\dagger}_Ld_L d^{\dagger}_Rd_R  +t_C (d^{\dagger}_Ld_R +H.c.)$ models
the coupled double QD, in which each QD has only a single level without
spin index. The purpose of neglecting the spin index is to emphasize that
fact that for this system, Kondo effect occurs without the need of
spin-degrees of freedom. Realistically, if a high magnetic field is applied 
to the QDs, 
leading the spin separation being larger 
than the QD energy level spacing\cite{ref6}, the opposite spin state 
essentially does not affect tunneling at low bias because its contribution
is actually weaker than that of the first excited QD state. In this
situation the spin-degrees of freedom can be neglected. Finally, when bias 
voltage $V$ is not very high, only one eigenstate on each QD is active, 
therefore the on-site interaction $U_o$ of each QD can be absorbed 
into its single particle level $\epsilon_{\alpha}$
($\alpha=L,R$),\cite{add2} 
and only the interdot off-site interaction $U$ will be considered.

Our theoretical analysis is from the Keldysh nonequilibrium Green's functions.
The current from the left lead flowing into the left QD can be expressed as
($\hbar=1$)\cite{ref10}: $I=-2e Im \int \frac{d\epsilon}{2\pi} \Gamma_L
\{ f_L(\epsilon) G^r_{LL}(\epsilon) + \frac{1}{2} G^<_{LL}(\epsilon)\}$,
where $f_{\alpha}(\epsilon)$ is the Fermi distribution of the $\alpha$ lead,
and $\Gamma_{\alpha}=2\pi \sum_k |t_{\alpha}|^2 \delta(\epsilon-\epsilon_k)$.
The Green's function $G^{r,<}_{LL}(\epsilon)$ are the Fourier transformation
of $G^{r,<}_{LL}(t)$, and 
$G^r_{LL}(t)\equiv -i\theta(t) <\{d_L(t), d^{\dagger}_L(0)\}>$,
$G^<_{LL}(t)\equiv i <d^{\dagger}_L(0)d_L(t)>$. Using the standard equation
of motion technique and taking the familiar decoupling
approximation\cite{ref11}, we have solved $G^r(\epsilon)$ in the infinite
$U$ limit ($U\rightarrow \infty$) to be:
\begin{equation}
  \left( \begin{array}{ll} 
      \epsilon-\epsilon_L - \Sigma^{(0)}_{L} -\Sigma^c_{L} -\Sigma^b_{R} &
         -t_C-\Sigma^d  \\   
         -t_C-\Sigma^d  &
      \epsilon-\epsilon_R - \Sigma^{(0)}_{R} -\Sigma^c_{R} -\Sigma^b_{L} 
         \end{array}
   \right) 
 \left( \begin{array}{ll} 
       G_{LL}^r &  G_{LR}^r \\
       G_{RL}^r &  G_{RR}^r
         \end{array}
   \right)   =
 \left( \begin{array}{ll}
       1-n_R & <d_R^{\dagger} d_L> \\
       <d_L^{\dagger} d_R>  & 1-n_L
         \end{array}
   \right)
\end{equation}
where $\Sigma^d=\Sigma^d_L+\Sigma^d_R$.
$\Sigma^{(0)}_{\alpha}(\epsilon)=\sum_k
|t_{\alpha}|^2/(\epsilon-\epsilon_{\alpha k}+i0^+)$ is the lowest-order
self-energy for a noninteraction system; the higher-order self-energy 
$\Sigma^b_{\alpha}$, $\Sigma^c_{\alpha}$, $\Sigma^d_{\alpha}$ are:
$
\Sigma^b_{\alpha}(\epsilon)=
 \sum_k [(\epsilon-\epsilon_{\alpha k})
(\epsilon\mp\Delta\epsilon -\epsilon_{\alpha k}) -2|t_C|^2] 
B_{\alpha k}(\epsilon)$,
$
\Sigma^c_{\alpha}(\epsilon)= \sum_k 2|t_C|^2
B_{\alpha k}(\epsilon) $,
and $
\Sigma^d_{\alpha}(\epsilon)=
\sum_k (\epsilon\mp\Delta\epsilon -\epsilon_{\alpha k})
t_C B_{\alpha k}(\epsilon)$,
with $B_{\alpha k}(\epsilon)= |t_{\alpha}|^2 f_{\alpha}(\epsilon_{\alpha k})
/ (\epsilon-\epsilon_{\alpha k})
/[(\epsilon-\Delta\epsilon -\epsilon_{\alpha k})
(\epsilon+\Delta\epsilon -\epsilon_{\alpha k}) -4|t_C|^2]$ and 
$\Delta\epsilon=\epsilon_L-\epsilon_R$. 
The quantity 
$n_{\alpha}=<d^{\dagger}_{\alpha} d_{\alpha}>$ in Eq.(1) is the intradot
electron occupation number, $<d_L^{\dagger}d_R>$ and $<d_R^{\dagger}d_L>$
are the interdot correlation due to the interdot tunnel coupling and the
interaction. They must be calculated self-consistently:
$<d^{\dagger}_{\alpha} d_{\beta}>=
-i\int \frac{d\epsilon}{2\pi} G^<_{\beta\alpha}(\epsilon)$. It should be
mentioned that although the equation of motion method only gives a correct
qualitative physics\cite{ref12}, it is sufficient for the purpose here. 

Next we solve the Keldysh Green's function $G^<(\epsilon)$. For interacting 
systems, $G^<(\epsilon)$ cannot be obtained from equation of motion without 
introducing additional assumptions\cite{ref12,ref13}. In previous work, 
various approximations were developed including the use of an ansatz for 
interacting lesser and greater self-energy\cite{ref13}; the noncrossing 
approximation\cite{ref12}, and so on. Any of them could be sufficient for 
our purposes. However, we note that since only $\int d\epsilon G^<(\epsilon)$ 
is actually needed for calculating both the electric current and for iterating
the self-consistent equation, one does not need to solve $G^<(\epsilon)$ 
itself. We found that quantity $\int d\epsilon G^<(\epsilon)$ can actually 
be solved {\it exactly}, allowing us to bypass any approximation involved in 
computing $G^<(\epsilon)$. We now present this technical advance which
is quite general and can be applied to many other systems. 
As stated above, $\int d\epsilon G^<_{\beta\alpha}(\epsilon) \sim
<d^{\dagger}_{\alpha}
d_{\beta}>$, we therefore proceed to first derive the operator
$d^{\dagger}_{\alpha}(t)d_{\beta}(t)$ ($\alpha,\beta=L,R$) over time $t$,
and then take average 
$<\frac{d}{dt} [d^{\dagger}_{\alpha}(t)d_{\beta}(t)]>=0$.
The last equality is true because we consider steady state transport. 
On the other hand, the time derivative can be explicitly calculated using
the Heisenberg equation, and we obtain
\begin{equation}
  (\epsilon_{\bar{\alpha}}-\epsilon_{\alpha} -i\frac{\Gamma_L+\Gamma_R}{2}
  ) <d^{\dagger}_{\alpha}d_{\bar{\alpha}}>
 +t_C<d^{\dagger}_{\alpha}d_{\alpha}>
 -t_C<d^{\dagger}_{\bar{\alpha}}d_{\bar{\alpha}}>
 =\int \frac{d\epsilon}{2\pi} 
            \Gamma_{\alpha}    f_{\alpha} 
          G^r_{\bar{\alpha}\alpha}
   - \int \frac{d\epsilon}{2\pi} 
       \Gamma_{\bar{\alpha}}  f_{\bar{\alpha}}
         G^a_{\bar{\alpha}\alpha}
\label{hei1}
\end{equation}
\begin{equation}
-t_C<d^{\dagger}_{\alpha}d_{\bar{\alpha}}>
 +t_C<d^{\dagger}_{\bar{\alpha}}d_{\alpha}>
-i\Gamma_{\bar{\alpha}}<d^{\dagger}_{\bar{\alpha}}d_{\bar{\alpha}}>
= \int \frac{d\epsilon}{2\pi}
         \Gamma_{\bar{\alpha}}       f_{\bar{\alpha}}
 [G^r_{\bar{\alpha}\bar{\alpha}}- G^a_{\bar{\alpha}\bar{\alpha}}]         
\label{hei2}
\end{equation}
where $\bar{\alpha}=R$ for $\alpha=L$, or $\bar{\alpha}=L$ for $\alpha=R$. 
From Eqs.(\ref{hei1},\ref{hei2}) and having already solved $G^r(\epsilon)$,
$<d^{\dagger}_{\alpha}d_{\beta}>$ ({\it i.e.} 
$\int d\epsilon G^<_{\beta\alpha}(\epsilon)$) can be exactly solved and
current can now be obtained without further difficulty.
In our numerical evaluations, we assume square bands of width $2W$
and symmetric coupling barriers so that
$\Gamma_{L}(\epsilon)=\Gamma_{R}(\epsilon)=\Gamma\theta(W-|\epsilon|)$, 
with $W=1000\gg max(k_B T,eV,\Gamma)$. Here $V=\mu_L-\mu_R$ is the bias
voltage. We also fix $\Gamma=1$ as energy unit and let $\mu_L=-\mu_R$.

First, we investigate LDOS of the DQD device.\cite{ref14} 
Because the LDOS of the right dot is similar to that of the left
one, we only discuss the latter. Consider equilibrium situation (bias $V=0$) 
and the case where the two dots have equal level position
($\Delta\epsilon\equiv \epsilon_L-\epsilon_R=0$), the LDOS is shown as the 
right inset of Fig.1d in which various curves correspond to different values 
of the interdot tunnel coupling $t_C$. When there is no tunnel coupling 
($T_C=0$, thin solid line), besides a broad main peak at $\epsilon \sim -1.5$ 
(not shown) which corresponds to the renormalized left dot level $\epsilon_L$,
a sharp peak emerges at energy $\epsilon=\mu$ (now $\mu=\mu_L=\mu_R$). This 
peak has the typical Kondo characteristic: it only exists at low temperature 
and is pinned at $\epsilon=\mu$. Turning on the tunnel coupling ($t_C\neq 0$) 
between the two dots, the Kondo peak is split into two smaller peaks at 
$\epsilon=\mu\pm 2t_C$ and whose weight are $\frac{1}{2}$\cite{ref15}. 
Setting the off-site Coulomb energy $U=0$, the Kondo peaks disappear. Clearly, 
it is the interdot off-site Coulomb interaction $U$ that induces the Kondo 
peak. 

The origin of the Kondo peaks at $t_C=0$ is from an interesting co-tunneling
process which is shown in the left inset of Fig.1d. When 
$\epsilon_L,\epsilon_R<\mu_L,\mu_R<\epsilon_L+U,\epsilon_R+U$, the DQD has 
only one electron ({\it e.g.} it occupies the right dot) and the first order 
tunneling is blocked by the off-site interaction. However, due to Heisenberg 
uncertainty principle, higher-order co-tunneling processes consisting of two 
virtual tunneling events can still take place. In such a process, the electron
in the right dot tunnels to the Fermi level of the right lead, followed by an 
electron in the left lead at the Fermi level tunneling into the left dot, on 
a very short time scale $\sim\hbar/(\mu_R-\epsilon_R)$. This co-tunneling 
process, in effect, moved the original electron from the right dot to the 
left one. At low temperatures, a frequent exchange of electrons between the 
two dots occur through this process even though $t_C=0$, and a coherent 
superposition of all possible co-tunneling processes of this type produces 
the Kondo resonance discussed in the last paragraph in which a very narrow 
peak emerges at $\epsilon=\mu$ in the LDOS curve of left (right) dot.
We emphasize that this mechanism of inducing a Kondo resonance is
qualitatively different from the co-tunneling process that induces the 
conventional Kondo effect by the on-site interaction: in that case the local 
spin is flipped after the co-tunneling process.

Next, we discuss the non-equilibrium situation ($V\neq 0$) keeping
$\Delta\epsilon=0$. Fig.1d shows the left dot's LDOS. At
$t_C=0$, a narrow Kondo peak emerges (thin solid line). However, its
position is at $\epsilon=\mu_R$, not at $\mu_L$. In other words,
the Kondo peak of one dot occurs at the chemical potential of the lead of
the other dot! This is a very surprising result indeed. It is also 
qualitivally different from the Kondo resonance induced by intradot 
interaction $U_o$ (not the interdot $U$) where the peak is pinned to the 
chemical potential of the lead attached to the same dot\cite{ref7}. 
When $t_C\neq 0$, states $\epsilon_L$ and $\epsilon_R$ hyberdize
into two ``molecular'' states 
$\epsilon^{\pm}=\frac{\epsilon_L+\epsilon_R}{2}\pm\frac{\Delta E}{2}$ expanding
to the entire device at $\Delta\epsilon=0$\cite{ref16}, 
where $\Delta E=\sqrt{\Delta\epsilon^2 +4t_C^2}$. The corresponding 
left dot's LDOS are plotted in Fig.1d. 
The main features are: (i) The Kondo peak at $\epsilon=\mu_R$ when 
$t_C=0$ is now split into three peaks, at $\epsilon=\mu_R$ and 
$\mu_R\pm\Delta E$ with weights $\frac{1}{2}$ and $\frac{1}{4}$ 
respectively. (ii) A dip emerges at $\epsilon=\mu_L$ and two 
more peaks are at $\mu_L\pm\Delta E$. Most importantly, all these (five) 
peaks and the dip have Kondo characteristics: (1) they are pinned at 
$\mu_{L/R}$ and $\mu_{L/R}\pm\Delta E$; (2) they disappear at high 
temperature; on the other hand, at lower temperature they become higher/lower 
for the peak/dip (See Fig.2); (3) they all disappear when $U=0$.

In fact, the five Kondo peaks and the dip are results of many-body
correlation due to the off-site interaction $U$.
They originate from the co-tunneling processes between the molecular state
$\epsilon^{\pm}$ and the two leads which are shown in Fig.1a-c.
For example, Fig.1a describes the following co-tunneling 
process. An electron initially occupies $\epsilon^+$ and tunnels to the 
Fermi level of the right lead. At almost the same moment another electron in
the left lead, having energy $\mu_R-\Delta E$, tunnels into the {\it other}
molecular state $\epsilon^-$. At low temperatures, a coherent superposition
of many co-tunneling events of this type gives rise to the Kondo resonance 
at $\mu_R-\Delta E$ in the LDOS. Exactly the same way, the co-tunneling 
events shown in Fig.1c induce a Kondo peak at $\mu_R+\Delta E$. Fig.1b shows 
a different type co-tunneling process. Here an electron occupies 
level $\epsilon^{\pm}$ initially. It then tunnels to the Fermi level
of the right lead, closely followed by another electron in the left lead 
with energy $\mu_R$ which tunnels into same level $\epsilon^{\pm}$. 
This co-tunneling process does not induce electron exchange between the 
molecular states $\epsilon^{\pm}$, therefore it cannot induce a Kondo peak 
at the DOS of the molecular states. However, molecular states $\epsilon^{\pm}$
are linear combinations of the original dot states $\epsilon_L$ and 
$\epsilon_R$, so in this co-tunneling process, electrons may still exchange
between $\epsilon_L$ and $\epsilon_R$, which gives rise to a Kondo peak at
$\epsilon=\mu_R$ in the left dot's LDOS and a dip at $\epsilon=\mu_R$ in the
right dot's LDOS.

The peculiar Kondo resonances discussed so far are for equal dot states
level positions, {\it i.e.} for $\Delta\epsilon=\epsilon_L - \epsilon_R=0$. 
They however persist even when $\Delta\epsilon \neq 0$. As $\Delta \epsilon$ 
increases from zero, the molecular states $\epsilon^{\pm}$ gradually localize 
to one of the dots and they asymmetrically couple to the two leads. 
If $\Delta\epsilon>0$, $\epsilon ^{+/-}$ are coupled stronger to the 
left/right lead, and weaker to the other lead (Fig.1a-c). Therefore, 
the co-tunneling processes of Fig.(1a,1b) will be weakened while
that of Fig.1c enhanced. Correspondingly, in the left dot's LDOS, the Kondo 
peaks at $\mu_R +\Delta E$ is increased, and others will be reduced (see 
inset of Fig.2). In the limiting case of $t_C/\Delta\epsilon \rightarrow 0$, 
only the Kondo peak at $\mu_R+\Delta E$ survives in the left dot's LDOS. It 
is not difficult to find that approximately the weights and positions for 
each left dot's Kondo peak are: 
$\frac{1}{2} (1-\frac{\Delta\epsilon}{\Delta E}) 
-\frac{|t_C|^2}{\Delta E^2}$ for Kondo peak at $\mu_R-\Delta E$;
$\frac{2|t_C|^2}{\Delta E^2}$ for Kondo peak at $\mu_R$;
$\frac{1}{2} (1+\frac{\Delta\epsilon}{\Delta E})
-\frac{|t_C|^2}{\Delta E^2}$ for Kondo peak at $\mu_R+\Delta E$;
$\frac{|t_C|^2}{\Delta E^2}$ for Kondo peak at $\mu_L-\Delta E$;
$\frac{-2|t_C|^2}{\Delta E^2}$ for Kondo dip at $\mu_L$; and
$\frac{|t_C|^2}{\Delta E^2}$ for Kondo peak at $\mu_L+\Delta E$.

The current $I$ and differential conductance $dI/dV$ are substantially
influenced by the Kondo resonances. The left inset of Fig.3 show $dI/dV$ 
versus $V$ for several different interdot coupling $t_C$. Since we fixed 
$\Delta\epsilon=0$, $dI/dV$ must be symmetric while $I$ must be antisymmetric 
across zero bias which is what we found. The main characteristics are: 
(1) A very narrow peak is exhibited at $V=0$ in $dI/dV$, as a result of
the co-tunneling process of Fig.1b. Increasing $t_C$, the height of this 
peak first increases sharply and then reaches a maximum $\sim 0.6 e^2/h$ (the
unitary value is $e^2/h$) at $t_C \sim 0.18\Gamma$, followed by a slow 
decrease. (2) $dI/dV$ abruptly drops at $V=\pm\Delta E$. If $t_C$ is large,
$dI/dV$ exhibits another peak at $V=\pm\Delta E$ followed by the drop.
Because at  $V=\pm\Delta E$, the left lead's chemical potential 
$\mu_L$ aligns with the left dot's Kondo resonance at 
$\mu_R\pm\Delta E$, while $\mu_R$ aligns with the right dot's Kondo resonance
at $\mu_L\mp\Delta E$. 
(3) Increasing temperature, 
the $dI/dV$ features at $V=0$ and $V=\pm\Delta E$ gradually diminish, and 
the overall value of $dI/dV$ greatly reduces (see the right inset of Fig.3). 
At high temperature, {\it e.g.} $T=0.5$, $dI/dV$ becomes very small and very
weakly dependent on bias.
 
When $\Delta \epsilon \neq 0$, $dI/dV$ is no longer perfectly symmetric 
as shown in Fig.3. With increasing 
$\Delta \epsilon$, electron tunneling between the left and right dots 
becomes more difficult and molecular states $\epsilon^{\pm}$ will 
localize to one of the dots. As a consequence, the features of $dI/dV$ 
at $V=0$ and $\pm\Delta E$ are reduced. Specifically, for 
$\Delta\epsilon>0$ the abrupt drop of $dI/dV$ at $V=-\Delta E$ will 
gradually disappear, because the co-tunneling process of Fig.1a becomes 
weaker. On the other hand, the abrupt drop at $V=\Delta E$ survives
although it becomes weaker. Furthermore, as discussed above, 
$\Delta\epsilon>0$ enhances the co-tunneling process of Fig.1c,
due to this enhancement a new peak in $dI/dV$ arises at $V=\Delta E$
for large $\Delta\epsilon$.  

In summary, we investigated Kondo effect in a DQD device with an off-site
electron correlation provided by the interdot interaction $U$. A number
of co-tunneling processes have been identified between the molecular states 
and the leads, and electrons may be moved from one dot to the other due to
these processes. Rather than flipping the electron spin, the interdot
exchange of electrons together with the off-site electron correlation and
coherent superpositions of the co-tunneling processes, induce several Kondo 
resonances without the electron spin-degrees of freedom.  We note that although 
the intradot interaction $U_o$ should be larger than $U$, the Kondo 
temperature is not monotonic in these parameters. It is actually quite 
probable that the Kondo resonance determined by $U$ is more prominent 
than that induced by $U_o$. Therefore, if spin, intradot and interdot
interactions are all present, more complicated Kondo phenomenon
should result.

{\bf Acknowledgments:}
We gratefully acknowledge financial support from NSERC of Canada and FCAR of
Qu\'ebec.


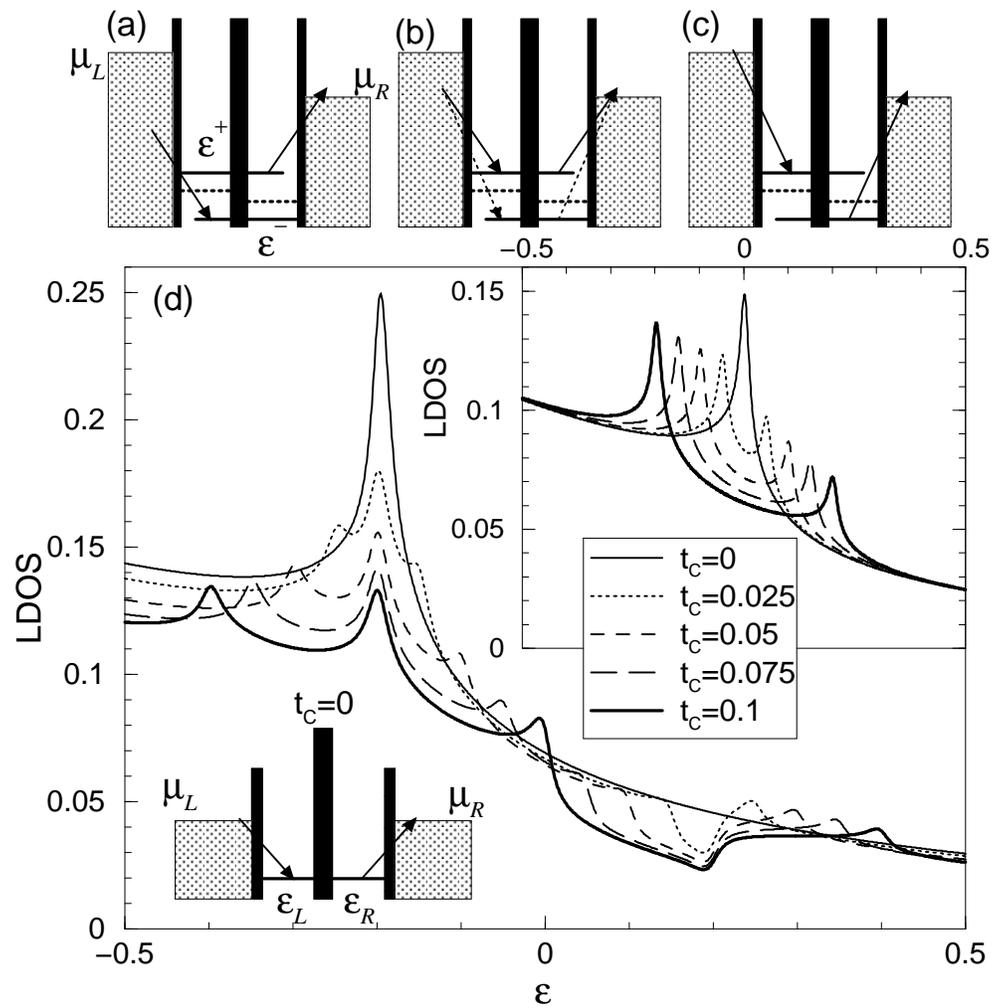
\begin{figure}
\caption{
(a)-(c) are schematic diagrams of co-tunneling processes between the two molecular
states and the two leads. (d) and its right inset show the left dot's LDOS vs
energy $\epsilon$ at different tunnel coupling $t_C$ with $\epsilon_L=\epsilon_R=-2.5$, 
$T=0.005$. In (d), $\mu_L=-\mu_R=0.2$; in its right inset
$\mu_L=-\mu_R=0$. The left inset of (d) is a schematic diagram
for the co-tunneling process at $t_C=0$.
}
\label{fig1}
\end{figure}

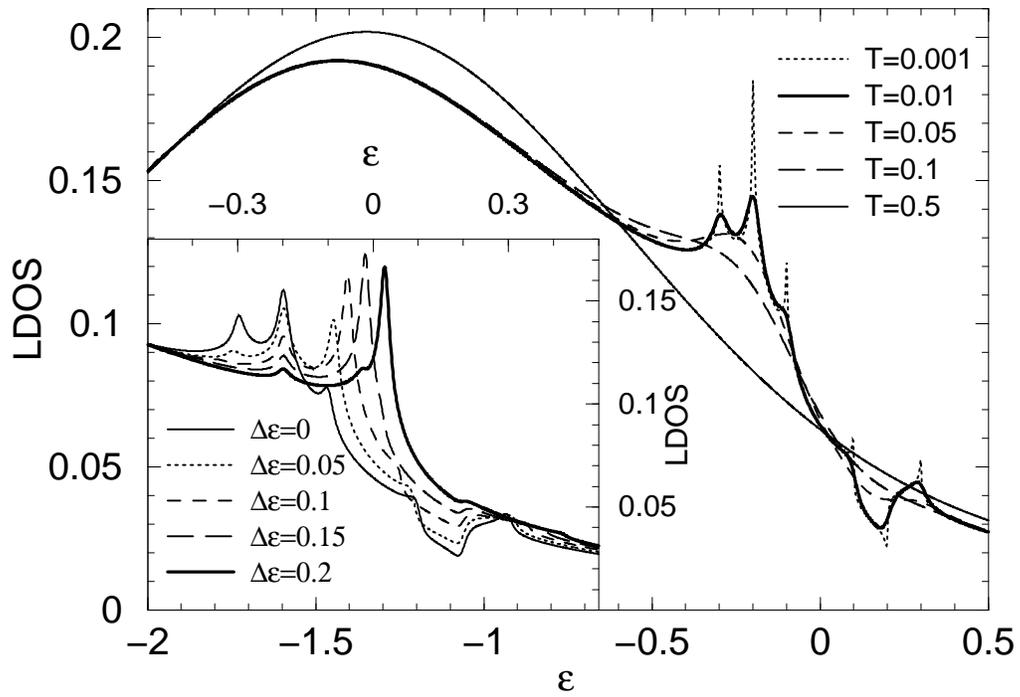
\begin{figure}
\caption{
Left dot's LDOS vs $\epsilon$ at different temperature $T$.
Inset: data for different $\Delta\epsilon$ and 
$\epsilon_L/\epsilon_R=-2.5 \pm \Delta\epsilon/2$. 
$t_C=0.05$ and other parameters are the same as those of Fig.(1d).
}
\label{fig2}
\end{figure}

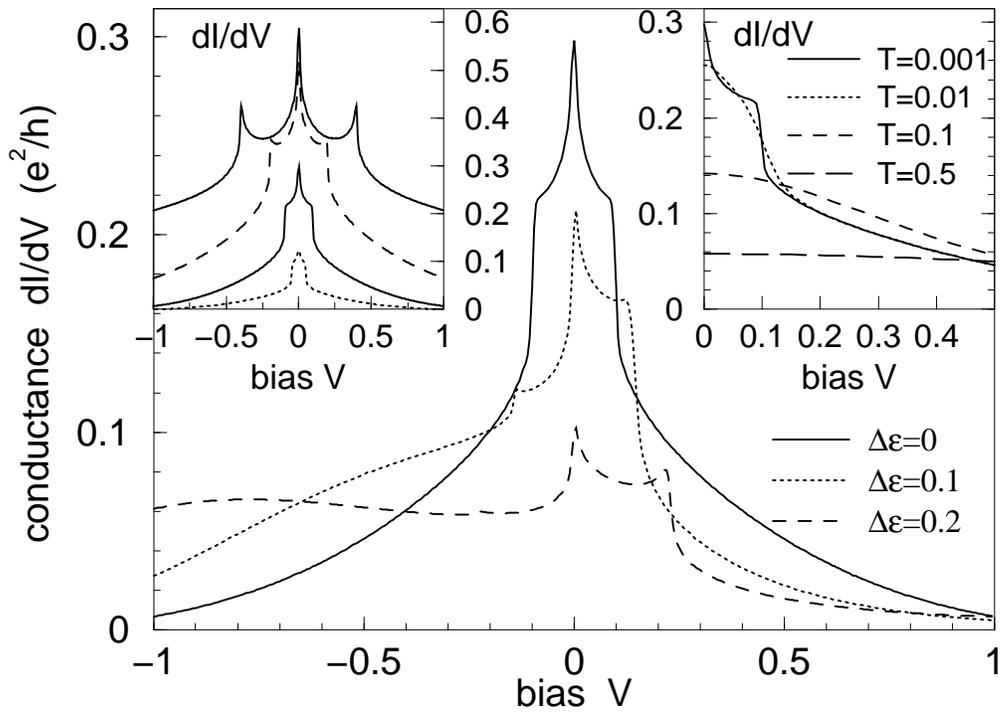
\begin{figure} 
\caption{ 
The conductance $dI/dV$ vs the bias $V$ at different $\Delta \epsilon$. 
$\epsilon_L=-2$, $\epsilon_R=-2-\Delta\epsilon$, $T=0.001$, and $t_C=0.05$.
Left inset: $dI/dV$ vs $V$ at different $t_C$ with $\epsilon_L=\epsilon_R=-2$ 
and $T=0.001$. Different curves correspond to $t_C=0.025$, $0.05$, $0.1$, 
and $0.2$ from bottom to top. Right inset: $dI/dV$ vs $V$ at different 
temperature $T$ where $t_C=0.05$ and $\epsilon_L=\epsilon_R=-2$.
} 
\label{fig3} 
\end{figure}


\begin{references}
\bibitem{ref11}
In our analysis, the decoupling approximation is:  
$ <\{Y X_1 X_2,d^{\dagger}_{\alpha'}\}> = 
<X_1 X_2> <\{ Y,d^{\dagger}_{\alpha'}\}>$, where
$X_{1/2}$ represent the operator $a_{\alpha k}$ and $a_{\alpha k}^{\dagger}$,
$Y$ is $d_{\alpha}$ and $d_{\alpha}^{\dagger}$.
\bibitem{ref15}
The Kondo peak's height is dependent on the weight and the Kondo
temperature. So two same weight Kondo peak, their height can be
difference.
\end{references}
\end{document}